\documentstyle[prl,aps,epsfig]{revtex}

\begin{document}


\title{ Indistinguishable multiplier statistics \\
        of \\
        discrete and continuous turbulent cascade models}

\author{
Bruno Jouault$^{1}$,
J\"urgen Schmiegel$^{1}$,
and
Martin Greiner$^{1,2,3}$
}

\address{$^1$Max-Planck-Institut f\"ur Physik komplexer Systeme, 
             N\"othnitzer Str.\ 38, D--01187 Dresden, Germany }
\address{$^2$Institut f\"ur Theoretische Physik, Technische Universit\"at,
             D--01062 Dresden, Germany }
\address{$^3$Department of Physics and Astronomy, Louisiana State
             University, Baton Rouge, LA 70803-4001, USA}

\date{22.09.1999}

\maketitle

\begin{abstract}
The multiplier statistics of discrete and continuous nonconservative
multiplicative cascade models, employed to describe the energy cascade
in fully developed turbulence, is investigated. It is found to be
indistinguishable due to small-scale resummation and restoration of
spatial homogeneity. Only the subclass of positively skewed weight
distributions, associated to the cascade generator, qualifies to
describe observed unconditional as well as conditional multiplier
distributions qualitatively; popular log-Poisson and log-stable
weight distributions do not share this property.
\end{abstract}

\newpage
\section{Introduction}

The intermittent small-scale dynamics of fully developed turbulence at 
very large
Reynolds numbers is believed to be more or less scale-invariant within 
the so-called inertial range, the latter being bound by the integral
length scale $L$ and the dissipation length scale $\eta \ll L$. 
Consequently, both experimental \cite{SRE97} and theoretical 
\cite{NEL94,FRI95} efforts largely concentrate on 
scaling exponents, which are deduced from structure functions of the 
velocity field. However, ``scaling exponents are not everything'' as
they represent for sure the simplest, but at the same time also the
most insensitive observables characterising multi-scale processes. 
Because of this insensitiveness any heroic effort to derive the 
observed multi-scaling directly from the Navier-Stokes equation
\cite{YAK98,BEL98} remains somewhat incomplete as some physics might
have been lost with the applied approximations and assumptions. Also,
due to the finiteness of the real-world inertial range and the limited
statistical sampling, the experimental procedure to extract 
multi-scaling
is not unequivocal and quoted values for scaling exponents beyond sixth
order should not be overemphasised. For these reasons and also in view 
of the quest for universality in the up-to-now unreachable limit 
$R_\lambda\rightarrow\infty$, additional observables other than 
scaling exponents are needed.

In order to explain this philosophy, ``scaling exponents are not 
everything'', in more detail, we discuss heuristic multiplicative
cascade processes and show that with a clever subclass of additional 
observables more
can be learned about the intrinsic cascade dynamics than only from
scaling exponents. Following Richardsons historic metaphor of
large eddies decaying into successively smaller and smaller eddies,
a multiplicative cascade process phenomenologically describes the 
turbulent redistribution of energy flux from large to small scales and
is able to reproduce the observed scaling exponents of the surrogate
energy dissipation field \cite{FRI95,MEN91}, which are related to those
of the velocity structure functions by the Refined Similarity 
Hypothesis \cite{STO94}. Originally intended to directly access the 
multiplicative 
weight distribution associated to the energy cascade generator 
distributions of so-called multipliers or break-up coefficients 
\cite{NOV71} have been extracted from very large Taylor-scale based
Reynolds number atmospheric boundary layer 
and wind tunnel flows recently \cite{CHH92,SRE95,PED96};
see also Ref.\ \cite{MOL95}.  
It was found that within the upper part $0 \ll l < L$ of the inertial 
range these multiplier distributions do in fact become 
scale-independent. However, they do depend on the relative 
position of parent and daughter 
domain; this non-homogeneity of the breakup leaves a subsequent
extraction of scaling exponents inconsistent \cite{NEL96} and
indicates the presence of correlations between successive multipliers, 
as has been confirmed by studying multiplier distributions conditioned 
on the value of the (scale-) previous multiplier \cite{SRE95,PED96}. 
The observed correlations appear to be in conflict with the simple 
multiplicative cascade models, where the cascade generator is assumed 
to be identical and independent at each breakup.
As has been convincingly demonstrated in Refs.\ \cite{JOU99a,JOU99b}
this apparent conflict can be resolved: introducing non-conservative
cascade generators with a positively skewed weight distribution
and restoring Euclidean homogeneity of the ultrametric cascade models, 
the multiplier distributions resulting from a discrete binary
multiplicative cascade become a scale-independent fix-point due to 
small-scale resummation and are in good qualitative agreement with 
the experimentally deduced distributions, including all observed
correlation effects.

Originally not anticipated, these multiplier distributions represent 
one of those wanted clever subclasses of additional observables, from which 
more can be learned about the relevance of cascade processes in turbulence
and about the intrinsic cascade dynamics than only from scaling exponents.
The experimental analysis \cite{SRE95,PED96} reveals that unconditional
multiplier distributions are observed to be scale-independent within the
range $500\eta \leq l \leq L/2$. In view of the findings of Ref.\ 
\cite{JOU99b} we call this the visible cascade range. Since due to 
small-scale resummation it takes about three binary scale steps for
the multiplier distributions to converge to the scale-independent fix-point
the lowest part of the true cascade range 
$\eta^* \approx 500\eta/2^3 \approx 60\eta \leq l \leq L$ 
is not visible. A further conclusion from this line of reasoning would be
that below $\eta^*$ dissipation dynamics sets in and modifies the 
scale-independent cascade dynamics. Besides this interpretation about the 
scope of cascade processes in turbulence, also more details about the 
intrinsic cascade dynamics can be learned from multiplier distributions
\cite{JOU99a,JOU99b}. Not every weight distribution, associated to a binary
cascade generator and reproducing observed scaling exponents within
experimental error bars, qualifies to yield the correct distributions of
multipliers and their correlations. It has to come with a positive 
skewness. Weight distributions of log-normal or certain asymmetric 
binomial type share this property, whereas, for example, a log-Poisson 
weight distribution \cite{SHE94} does not.

All those conclusions have been drawn from binary discrete multiplicative
cascade processes in connection with base-two ($\lambda=2$)
multiplier distributions. Multiplier distributions referring to other 
scale steps have not been looked at. In this respect it is of interest 
to find out whether the latter do again reveal a scale-independent 
fix-point behaviour with additional correlations or whether they are
simply artefacts of the binary discrete cascade model implementation.
Not only this aspect is of relevance to find answers, the impact of
cascade model implementations with scale steps other than 
$\lambda^\prime = 2$, i.e.\ $\lambda^\prime > 2$ or 
$\lambda^\prime \rightarrow 1$, has to be studied for the
multiplier phenomenology: do multiplier distributions resulting from
different model implementations differ or not? Put in other words: 
are different cascade model implementations distinguishable or 
indistinguishable on the level of multiplier observables?
--
Theoretically the scale step $\lambda^\prime$ associated to the cascade 
generator is not fixed. Due to the present lack of a derivation of 
cascade models from the Navier-Stokes equation, no specific choice is 
favoured for $\lambda^\prime$, except for personal taste. 
Only if the last question can be answered with yes, we can consider
multiplier distributions as a clever subclass of additional observables.

The organisation of the paper is as follows: In Sect.\ II we 
sketch the implementation of general multiplicative cascade processes
and briefly summarize previous work on ($\lambda = 2$)-multiplier 
distributions resulting from a discrete ($\lambda^\prime = 2$) model
implementation. Sect.\ III concentrates on discrete 
($\lambda^\prime \geq 2$) implementations
of multiplicative cascade processes and presents multiplier distributions 
referring to scale steps $\lambda \neq 2$. A quasicontinuous implementation
of random multiplicative cascade models is used in Sect.\ IV to study
respective multiplier distributions referring to various scale steps.
So-called log-stable weight distributions \cite{SCH87,SCH92a}
are considered in Sect.\ V and, although leading to correct 
scaling exponents \cite{KID91,SCH92b}, shown not to reproduce the 
observed multiplier distributions. Several conclusions are presented 
in Sect.\ VI, the most important being that it does not matter how 
multiplicative cascade models are implemented (discrete or continuous), 
since the multiplier observables are indistinguishable.

\section{Implementation of discrete and continuous
         multiplicative cascade models}

\subsection{Implementation}

The following one-dimensional implementation of multiplicative cascade
processes is designed to model the energy cascade from large ($L$) to
small ($\eta^*$) length scales:

The initial energy flux density field at the integral scale $L$ is
set equal to $\varepsilon (x,L) = 1$ without any loss of generality.
For an intermediate length scale $\eta^* \leq l < L$ we introduce a
so-called weight field
\begin{equation}
\label{two1}
  q(x,l)  
    =  \sum_{k=-\infty}^{\infty} q_k(l) 
       \Theta (x - k l , l )
       \; ,
\end{equation}
where the index function
\begin{equation}
\label{two2}
  \Theta ( x , l )
    =  \left\{
       \begin{array}{ll}
         1  &  \qquad (0 \leq x < l) \\
         0  &  \qquad (x < 0, \, x \geq l)
       \end{array}
       \right.
\end{equation}
represents a subdomain of length $l$. Each random weight $q_k(l)$ 
is independently
drawn from a scale-independent probability distribution $p(q)$ with 
mean $\langle q \rangle = 1$. The energy density field 
\begin{equation}
\label{two3}
  \varepsilon (x,\eta^*)
    =  \left( \prod_{i=1}^n q(x,l_i) \right) \varepsilon (x,L)
\end{equation}
at the smallest scale $\eta^*$ is then constructed as a 
multiplication of weight fields belonging to a hierarchy of
different intermediate length scales $l_i = L/(\lambda^\prime)^i$, where
$(\lambda^\prime)^n = L/\eta^*$.

Let us discuss the special case for $\lambda^\prime=2$, so that it
is sufficient to consider the fields only in the domain $0 \leq x < L$.
The implementation (\ref{two3}) can then be understood in terms of a binary
cascade generator: The energy density $\varepsilon(l)$ of a domain of 
length $l$ is redistributed onto a left and right subdomain of length $l/2$
with energy densities $\varepsilon_L(l/2) = q_L \varepsilon(l)$ and 
$\varepsilon_R(l/2) = q_R \varepsilon(l)$, where the random weights are
drawn from the splitting function 
\begin{equation}
\label{two4}
  p(q_L,q_R) = p(q_L) p(q_R)  \; .
\end{equation}
The successive application of this cascade generator from the large scale 
$L$ down to the small scale $\eta^*$ leads to the field (\ref{two3}).
--
The splitting function (\ref{two4}) is not in its most general form as it
does not need to factorise. However, it has been noted in Ref.\ 
\cite{JOU99b}, that because fully developed turbulence is a $3+1$ dimensional
process, the observational reduction of a more-dimensional multiplicative
cascade process to one dimension leads to a splitting function which
factorises almost completely.

\subsection{Observables}

The standard observable of multiplicative cascade processes are scaling
exponents representing the intermittency corrections.
They are deduced from a scale-invariant moment analysis
of the synergetic field (\ref{two3}); a naive derivation gives
\begin{equation}
\label{two5}
  \left\langle \varepsilon(x,l_m)^\nu \right\rangle
    =  \left\langle q(x,l)^\nu \right\rangle^m
    =  \left( \frac{L}{l_m} \right)^{\tau(\nu)}
    =  (\lambda^\prime)^{m \tau(\nu)}
\end{equation}
with scaling exponents
\begin{equation}
\label{two6}
  \tau(\nu)
    =  \frac{\ln\langle q^\nu \rangle}{\ln\lambda^\prime}
       \; ,
\end{equation}
where the independence of the $m$ weights $q(x,l_1)$, \ldots , $q(x,l_m)$
has been exploited. This deduction is not in full accordance with the analysis
of experimentally measured fields; those are recorded at the finest
resolution scale $\eta < \eta^*$, then averaged over 
larger length scales $l$,
\begin{equation}
\label{two7}
  \overline{\varepsilon}(x,l)
    =  \frac{1}{l}
       \int_{x-l/2}^{x+l/2} \varepsilon(x^\prime,\eta) dx^\prime
       \; ,
\end{equation}
and are not identical to the model field evolved from scales $L$ to $l$
\cite{JOU99b,SCH87}, i.e.\
\begin{equation}
\label{two8}
  \overline{\varepsilon}(x,l)
    \neq  \varepsilon(x,l)
          \; .
\end{equation}
The scale dependence of moments 
$\langle \overline{\varepsilon}(x,l)^\nu \rangle$
of the backward field are different from
(\ref{two5}) and do not show rigorous scaling at the very small
scales \cite{GRE96}. Only in the large-scale asymptotic scaling
exponents can be extracted unambiguously and coincide with
the expression (\ref{two6}).

Originally proposed in Ref.\ \cite{NOV71} and applied in Refs.\
\cite{CHH92,SRE95,PED96,MOL95} the method of random multipliers (or
break-up coefficients) aims at directly accessing scaling exponents
via respective distributions of the former. 
In its most general form multipliers are defined as
\begin{equation}
\label{two9}
  M(x,l;\lambda,\Delta)
    =  \frac{\overline{\varepsilon}(x^\prime,\frac{l}{\lambda})}
            {\lambda \overline{\varepsilon}(x,l)}
       \; ,
\end{equation}
where 
$\overline{\varepsilon}(x^\prime,\frac{l}{\lambda})$
represents the averaged energy density over an offspring interval of
length $l/\lambda$ and with its centre shifted by 
$x^\prime - x = ( 1 - \lambda^{-1} ) l\Delta$
with respect to the parent interval. We distinguish three special cases
for the shift parameter $\Delta$: 
for $\Delta=0$, $\pm 1/2$ we call $M(x,l;\lambda,\Delta)$
centred, right- and left-sided, respectively.
--
For very large Reynolds number turbulent flows it has been found
experimentally \cite{CHH92,SRE95,PED96} that the unconditional
multiplier distributions $p(M(l;\lambda,\Delta))$, which are sampled over
$x$, do not depend on the scale $l$ within the regime 
$\eta \ll l \leq L/2$. However, a dependence on the shift parameter
$\Delta$ was noticed; this ``nonhomogeneity of the breakup'' 
obscures the relationship between multiplier and weight distribution 
and the once intended extraction of scaling exponents \cite{NEL96}.
Actually the $\Delta$-dependence hints to the existence of
correlations between successive multipliers.
These correlations have been observed in conditional multiplier distributions
$p ( M(x^\prime, l/\lambda;\lambda,\Delta) | M(x,l;\lambda,\Delta) )$
with left/right-sided ($\Delta=\pm 1/2$) \cite{SRE95}
and centred ($\Delta=0$) \cite{PED96} shift parameter, respectively, and
with the emphasis on base-two ($\lambda=2$) scale steps. 

These latest findings appear to contradict the simple multiplicative cascade
processes with independent breakups. However, this statement is not true!
The studies in Refs.\ \cite{JOU99a,JOU99b} with nonconservative binary
($\lambda^\prime=2$) random multiplicative cascade models have given a new
interpretation to multiplier distributions:
unconditional multiplier distributions represent a scale-independent
fix-point due to small-scale resummation; the ``nonhomogeneity of the
breakup'' is also naturally explained along this line. Once the sampling
of multipliers is allowed over all and not only dyadic positions the
correct multiplier correlations have been obtained from positively
skewed weight distributions; for example, a log-normal weight
distribution does qualify whereas a log-Poisson distribution does not
qualify. These results show that multiplier distributions do reveal more
information about the underlying weight distribution than might be
extracted from approximate scaling exponents.

\section{Multiplier statistics of discrete multiplicative cascade models}

\subsection{Discrete multiplicative cascade models with binary splittings}

For discrete multiplicative cascade models with binary splittings the 
scale step entering into the construction of the energy density field
(\ref{two3}) is set equal to $\lambda^\prime=2$. 
Furthermore, we choose $n=10$. The length of the field configuration 
is set equal to $0 \leq x \leq 10^5 L$, 
mimicking a long time series. As weight distribution we choose the 
binomial distribution  
\begin{equation}
\label{three1}
  p(q)
    =  \frac{\alpha_2}{\alpha_1+\alpha_2} \delta\left( 1-(1-\alpha_1) \right)
       + \frac{\alpha_1}{\alpha_1+\alpha_2} \delta\left( 1-(1+\alpha_2) \right)
\end{equation}
with parameters $\alpha_1=0.3$, $\alpha_2=0.65$; 
this choice has the correct skewness
$S_3 = \langle ( q - \langle q \rangle )^3 \rangle / 
 ( \langle ( q - \langle q \rangle )^2 \rangle )^{3/2}
 = (\alpha_2-\alpha_1)/\sqrt{\alpha_1\alpha_2} > 0$
in order to reproduce conditional base-two multiplier distributions 
\cite{JOU99a}. Other, non-binomial weight distributions, such as of
log-normal or Pearson-III type, are also able to provide a positive
skewness and lead to more or less identical base-two multiplier
distributions \cite{JOU99b}, all being in 
qualitative accordance with experimental
findings \cite{SRE95,PED96}. Also multiplier distributions
different from base-two do not distinguish the aforementioned
weight distributions. Hence, we will only
show results for the weight distribution (\ref{three1}).

Due to small-scale resummation \cite{JOU99b}
the base-two multiplier distributions are
scale-invariant in the regime $8\eta^* \leq l \leq L/2$;
for $\eta^*\leq l\leq 8\eta^*$ the distributions converge to the 
scale-independent fix-point and for scales very close to the integral
length scale deviations from scale-invariance set in because of the 
finiteness of $L$. Figure 1(a2) shows
simulation results for left-sided multiplier distributions
($\Delta = -0.5$) at scale $l=L/8$:
the unconditional multiplier distribution $p(M(l;2,-0.5))$
comes close to a Beta-distribution parametrisation
\begin{equation}
\label{three2}
  p(M)
    =  \frac{\Gamma(\beta_1+\beta_2)}{\Gamma(\beta_1)\Gamma(\beta_2)}
       M^{\beta_1-1} (1-M)^{\beta_2-1}
\end{equation}
with $\beta_1=\beta_2=3.2$, which has been extracted from 
very large Reynolds number atmospheric boundary layer turbulence
\cite{SRE95}. Once conditioned on the respective parent
multiplier, correlations are observed: 
the distribution $p(M(l;2,-0.5)|0.5 \leq M(2l;2,-0.5) \leq 1)$
with a large parent multiplier becomes broader than the unconditional 
distribution and its average is shifted to a value larger than 
$1/2$. The opposite holds for the conditional distribution
$p(M(l;2,-0.5)|0 \leq M(2l;2,-0.5) \leq 0.5)$ with a small parent 
multiplier; it is more narrow than the unconditional distribution and its
average is shifted to a value smaller than $1/2$. The positive correlation
between left-sided daughter and left-sided parent multiplier is a
consequence of the homogeneous sampling of multipliers
over all $x$-positions, which is not 
restricted to dyadic cascade positions. Respective multiplier distributions of
right/right combinations are identical to these left/left combinations. 
A negative correlation between daughter and parent multipliers is revealed
for identical left/right and right/left combinations; their respective 
distributions are mirror images of those for left/left and right/right
combinations, i.e.\ 
$p(M(l;2,0.5)|\underline{M} \leq M(2l;2,-0.5) \leq \overline{M}) 
 = p(1-M(l;2,-0.5)|\underline{M} \leq M(2l;2,-0.5) \leq \overline{M})$.
This explains that conditional distributions, which are averaged over all
four daughter/parent combinations, are again symmetric around 
$M(\lambda=2)=1/2$;
these distributions have been shown in Refs.\ \cite{SRE95,JOU99a,JOU99b}.

Left-sided multiplier distributions for scale steps different from
$\lambda=2$ are shown in Figs.\ 1(a1+a3). For $\lambda=4$ the
unconditional multiplier distribution converges to a fix-point for
$\eta^* \ll l \leq L/2$; its maximum is shifted towards smaller values
since its average amounts to 
$\langle M(\lambda=4) \rangle = 1/\lambda = 0.25$.
A conditioning on a left-sided parent multiplier again reveals correlations
between successive multipliers: the distribution 
$p( M(l;4,-0.5) | 0.25 \leq M(4l;4,-0.5) \leq 1 )$
is broader than the unconditional one, whereas
$p( M(l;4,-0.5) | 0 \leq M(4l;4,-0.5) \leq 0.25 )$
is more narrow.
--
Also multiplier distributions of base step close to one, e.g.\ 
$\lambda = 2^{1/4}$, show fix-point behaviour in the regime
$\eta^* \ll l \leq L/2$. In view of the discrete 
($\lambda^\prime=2$)-steps used
in the cascade evolution this result appears unexpected on first sight,
but again it is small-scale resummation, which explains this result on
second thought. The unconditional left-sided distribution comes with an
average of
$\langle M(\lambda=2^{1/4}) \rangle = 1/\lambda = 0.84$
close to one. The respective conditional distributions
$p( M(l;2^{1/4},-0.5) | 0 \leq M(2^{1/4}l;2^{1/4},-0.5) \leq 0.84 )$
and
$p( M(l;2^{1/4},-0.5) | 0.84 \leq M(2^{1/4}l;2^{1/4},-0.5) \leq 1 )$
both deviate from the unconditional distribution, but contrary to the
$\lambda = 2$ and $\lambda = 4$ cases, the former has now
become broader whereas the latter turns out to be more narrow. Closer
inspection reveals that this tendency has been reversed at around
$\lambda \approx 1.75$.

Centred multiplier distributions for scale steps $\lambda=4$, $2$ and 
$2^{1/4}$ do qualitatively show the same behaviour as the respective 
left-sided multiplier distributions depicted in Figs.\ 1(a1)-(a3). All 
centred distributions are more narrow than their respective left-sided 
counterparts, reflecting again the ``nonhomogeneity of the breakup''. 
This is also illuminated in Fig.\ 2, where the $\lambda$-dependence
of the two exponents $\beta_1$ and $\beta_2$ of the Beta-distribution
(\ref{three2}), fitted to the respective unconditional left-sided and 
centred multiplier distributions, are shown. Since according to 
(\ref{three2}) we have $\langle M \rangle = \beta_1/(\beta_1+\beta_2)$, 
the two exponents are related by setting 
$\langle M \rangle = \lambda^{-1}$.

Conditional multiplier distributions of base step different from 
$\lambda = 2$ have not been looked at in Refs.\ \cite{SRE95,PED96}.
Hence, we consider our respective simulation results as predictions.

\subsection{Discrete multiplicative cascade models with triple and
            quadruple splittings}

Most often discrete multiplicative cascade models are chosen with scale
steps $\lambda^\prime=2$. Now we will consider $\lambda^\prime=3$ and 
$\lambda^\prime=4$ and investigate the resulting multiplier distributions.
Within the implementation (\ref{two1})-(\ref{two3}) the parameters are 
chosen as $n=7$ and $0\leq x\leq 10^5L$. Again for 
demonstration, the binomial weight distribution (\ref{three1}) is picked
with new parameters $\alpha_1=0.37$, $\alpha_2=0.8$ for $\lambda^\prime=3$
and $\alpha_1=0.42$, $\alpha_2=0.92$ for $\lambda^\prime=4$. According to
the expression (\ref{two6}) the two parameter settings yield an acceptable
intermittency exponent $\tau(2)=0.236$, but actually have been fitted
to reproduce the experimentally deduced unconditional left-sided 
base-two ($\lambda=2$) multiplier distribution (\ref{three2}) 
with $\beta_1=\beta_2=3.2$ in the fix-point scale regime $\eta^*\ll l<L/2$.
For both cases the resulting unconditional and conditional left-sided 
fix-point multiplier distributions of bases $\lambda=4$, $\lambda=2$ and
$\lambda=2^{1/4}$ are almost
indistinguishable from the respective distributions
shown in Figs.\ 1(a1)-(a3) for the binary discrete multiplicative cascade
implementation with $\lambda^\prime=2$. The same holds for centred
multiplier distributions.

For parameter choices $\alpha_1$, $\alpha_2$, resulting in a negative 
skewness of the binomial weight distribution, the shape of the resulting
unconditional left-sided fix-point multiplier distribution turns out to 
deviate from the symmetric Beta-distribution parametrisation to some 
small, but noticeable extend. Moreover, the correct effects observed 
in the conditional multiplier distributions can not be reproduced. 
-- 
On the contrary, well adjusted log-normal weight distributions come again
with the correct positive skewness and almost identically reproduce all 
unconditional and conditional multiplier distributions derived from the
good binomial weight parametrisations.

\section{Multiplier statistics of continuous 
         multiplicative cascade models}

\subsection{Continuous multiplicative cascade models}

A discrete implementation of random multiplicative cascade processes is 
often opposed by the question "why discrete?". It is true that a 
continuous implementation leads to nicer mathematics in terms of 
log-infinite divisible and log-stable distributions 
\cite{SCH87,SCH92a,KID91,SCH92b,NOV94,SHE95}, 
but since the relationship of the
random multiplicative cascade processes to the Navier-Stokes equation
is unclear for the moment we prefer to consider discrete and continuous
versions on an equal footing. 
-- 
As far as the numerical implementation is concerned a continuous random 
multiplicative cascade process will always be quasi-continuous. Hence,
Eqs.\ (\ref{two1})-(\ref{two3}) do apply with a quasi-continuous
scale step $\lambda^\prime$ close to one; we choose 
$\lambda^\prime = 2^{1/8}$, which compared to a binary cascade corresponds
to an eightfold scale-densification. Other parameters are set equal to
$n=80$ and $0\leq x\leq 10^5L$. 
Due to scale-densification the binomial weight distribution (\ref{three1})
now comes with rescaled parameters. The setting $\alpha_1=0.1$, 
$\alpha_2=0.22$ comes from a fit of the resulting unconditional 
left-sided base-two multiplier distribution to the expression
(\ref{three2}) with $\beta_1=\beta_2=3.2$; see
Fig.\ 1(b2). Using Eq.\ (\ref{two6}) it also yields an intermittency
exponent of $\tau(2)=0.25$. 

Because of small-scale resummation all multiplier distributions become
again scale-independent fix-points in the range $\eta^*\ll l\leq L/2$; 
to be more concrete, the simulations reveal a lower bound of
approximately $8\eta^*$. The unconditional left-sided multiplier 
distributions of base-scale $\lambda=4$, $2$ and $2^{1/4}$, which are
shown in Figs.\ 1(b1)-(b3), are hard to distinguish from those obtained 
from a scale-discrete cascade implementation, which are illustrated in 
Figs.\ 1(a1)-(a3). A similar statement can be made not only about the
respective left-sided conditional distributions, but also for all 
unconditional and conditional centred multiplier distributions. 
This finding demonstrates that multiplier distributions in the fix-point
regime do not distinguish discrete and continuous versions
of random multiplicative cascade processes; again, the reason for this is
small-scale resummation.

Other suitable parameter settings of the binomial weight distribution
quite easily reproduce all unconditional multiplier distributions shown 
so far. Even with a different-signed skewness, as for example with 
$\alpha_1=0.195$, $\alpha_2=0.1$, the correlation effects observed in the
resulting conditional multiplier distributions go in the right direction,
but are reduced by a factor of about $2$ in magnitude. 
--
Another interesting parameter setting is 
$\alpha_1=1-\gamma_1(\lambda^\prime)^{\gamma_2}$,
$\alpha_2=(\lambda^\prime)^{\gamma_2}-1$. 
In the limit $\lambda^\prime \rightarrow 1^+$ with $\gamma_1=\gamma_2=2/3$
the weight distribution comes with a large negative skewness 
$S_3=-1/\sqrt{2(\lambda^\prime-1)}$ and
leads to the infinite divisible log-Poisson cascade model of Refs.\ 
\cite{SHE94,SHE95}. All 
unconditional multiplier distributions shown so far are again reproduced,
but only very weak correlation effects are observed in the resulting 
conditional multiplier distributions. This demonstrates that although
the log-Poisson cascade model is capable to reproduce scaling
exponents with no doubt, it fails to describe the correlation 
systematics of multiplier distributions.

These findings suggest that in order to reproduce unconditional as well as
correct conditional multiplier distributions the weight distribution has 
to possess a positive skewness, irrespective of the chosen 
scale-step-$\lambda^\prime$ implementation of the underlying multiplicative 
cascade process. For the binomial weight distribution (\ref{three1}) we note
that for the qualitatively best parameter choices, i.e.\ 
($\alpha_1$, $\alpha_2$) $=$ 
($0.1$, $0.22$), ($0.3$, $0.65$), ($0.37$, $0.8$), ($0.42$, $0.92$)
for $\lambda^\prime=2^{1/8}$, $2$, $3$, $4$, respectively,
the resulting skewness almost coincides at a value around 
$S_3 \approx 0.8$. Here a more quantitative analysis, driven by very large
Reynolds number data, would certainly deserve future consideration.
--
Multiplier distributions resulting from a meticulously tuned log-normal 
weight distribution 
\begin{equation}
\label{fourA1}
  p(q)  
    =  \frac{1}{\sqrt{2\pi}\sigma q}
       \exp\left( 
       - \frac{1}{2\sigma^2}
       \left( \ln q + \frac{\sigma^2}{2} \right)^2
       \right)  
\end{equation}
are also practically indistinguishable from the results 
shown in Fig.\ 1. Qualitative best parameters for the implementations
with $\lambda^\prime=2^{1/8}$, $2$, $3$, $4$ are
$\sigma=0.15$, $0.42$, $0.55$, $0.60$, respectively, reflecting an 
approximate, but obvious log-normal dependence 
$\sigma \sim \sqrt{\log_2\lambda^\prime}$. Contrary to the case of the
binomial weight distribution, the skewness of the respective log-normal 
distributions is not identical; it increases with increasing 
$\lambda^\prime$ and stays positive for all four cases.

\subsection{Continuous-turned-discrete multiplicative cascade models}

The multiplier statistics of continuously
implemented multiplicative cascade processes appears to be indistinguishable
from their discrete counterparts. This poses the question to what degree
a continuous multiplicative cascade process might be described by an
effective discrete multiplicative cascade process.

For the translation of a continuous into a discrete multiplicative
cascade process it is important to deal with the forward field. In analogy
to Eq.\ (\ref{two3}) the energy density field is evolved from integral
down to the intermediate binary target length scales $l$ and $l/2$, 
respectively, i.e.\
\begin{eqnarray}
\label{fourB1}
  \varepsilon(x,l)
    & = &  \left( \prod_{i=1}^{n_l} q(x,l_i) \right) \varepsilon(x,L)
           \; ,  \nonumber \\
  \varepsilon(x,\frac{l}{2})
    & = &  \left( \prod_{i=n_l+1}^{n_l+d} q(x,l_i) \right) \varepsilon(x,l)
\end{eqnarray}
with $2\eta^* \leq l \leq L$, $(\lambda^\prime)^{n_l} = L/l$ and the 
scale-densification $d=\ln 2/\ln\lambda^\prime$. 
We restrict the $x$-regime to 
$0 \leq x \leq 10^5L$. The averaging
\begin{eqnarray}
\label{fourB2}
  \underline{\varepsilon}(x,l)
    & = &  \frac{1}{l} \int_{x}^{x+l} \varepsilon(x^\prime,l) dx^\prime
           \; ,  \nonumber \\
  \underline{\varepsilon}_L(x,\frac{l}{2})
    & = &  \frac{2}{l} \int_{x}^{x+l/2} \varepsilon(x^\prime,l/2) dx^\prime
           \; , \\
  \underline{\varepsilon}_R(x,\frac{l}{2})
    & = &  \frac{2}{l} \int_{x+l/2}^{x+l} \varepsilon(x^\prime,l/2) dx^\prime
           \nonumber 
\end{eqnarray}
with discrete positions $x = k l$, $0 \leq k < 10^5L/l$ is principally
different from the backward averaging of Eq.\ (\ref{two7}) and leads to the
binary weights
\begin{eqnarray}
\label{fourB3}
  q_L
    & = &  \frac{\underline{\varepsilon}_L(x,\frac{l}{2})}
                {\underline{\varepsilon}(x,l)}
           \; ,  \nonumber \\
  q_R
    & = &  \frac{\underline{\varepsilon}_R(x,\frac{l}{2})}
                {\underline{\varepsilon}(x,l)}
           \; .
\end{eqnarray}
Sampling over all discrete positions leads to a binary splitting function 
$p_{\rm eff}(q_L,q_R)$, 
which still might depend on the scale $l$.

For the numerical simulations of a concrete quasi-continuous cascade 
process again an eightfold scale-densification ($d=8$) has been chosen 
together with the binomial weight distribution (\ref{three1})
with parameters $\alpha_1 = 0.1$, $\alpha_2 = 0.22$. The effective binary
splitting function $p_{\rm eff}(q_L,q_R)$
with effective weights (\ref{fourB3}) is illustrated in Fig.\ 3a for an
intermediate length scale. For $\eta^* \leq l \leq L/2$ it is practically 
scale-independent; only at the very large scales a weak scale-dependence
is observed. The projected weight distributions
$p_{\rm eff}(q_L) = \int p_{\rm eff}(q_L,q_R) dq_R$
and 
$p_{\rm eff}(q_R) = \int p_{\rm eff}(q_L,q_R) dq_L$
are identical and come close to a log-normal distribution with a positive
skewness.
Observe in Fig.\ 3b that the projected weight distribution is not simply 
a $d$-fold convolution of the binomial weight distribution (\ref{three1});
the modification is a result of the multivariate implementation (\ref{two1})
of the weight fields.
Note also, that the effective splitting function does not strictly
factorise; a small correlation between the two effective weights
$q_L$ and $q_R$ exists as the relative entropy
$S_{\rm rel} 
 = -\int dq_L dq_R p_{\rm eff}(q_L,q_R) 
   \ln( p_{\rm eff}(q_L,q_R) / p_{\rm eff}(q_L) p_{\rm eff}(q_R) )
 \approx -0.1
 < 0$
turns out to be negative.

So far the continously implemented multiplicative cascade process
has been directly translated into a ($\lambda^\prime=2$)-discrete 
multiplicative cascade process with an effective splitting function
$p_{\rm eff}(q_L,q_R)$. For consistency we still 
have to check, whether this effective 
($\lambda^\prime=2$)-discrete multiplicative cascade 
process leads to identical results for the multiplier statistics. The
discrete implementation (\ref{two1})-(\ref{two3}) with $\lambda^\prime=2$ 
and $p_{\rm eff}(q_L,q_R)$ 
instead of Eq.\ (\ref{two4}) leads to the multiplier distributions shown 
in Figs.\ 1(c1)-(c3). Qualitatively there is no difference to the 
distributions illustrated in Figs.\ 1(a1)-(a3), (b1)-(b3) and
quantitative differences remain very small.

\section{No room for log-stability}

In this Section we want to demonstrate again, that ``scaling exponents is
not everything''. In view of renormalisation theory so-called log-stable 
distributions represent mathematically very attractive parametrisations for
the weight distribution $p(q)$, leading its way to the denotation 
``universal multifractals'' \cite{SCH92a}. 
In connection with the energy cascade
in fully developed turbulence they have already been discussed in Refs.\
\cite{KID91,SCH92b}, where the expression
\begin{equation}
\label{five1}
  \tau(\nu)
    =  \tau(2) \frac{n^\alpha-n}{2^\alpha-2}
\end{equation}
has been derived for the scaling exponents (\ref{two6}). With an 
intermittency exponent of $\tau(2)=0.22$ and a stable index of
$\alpha\approx 1.6$ the experimentally deduced scaling exponents
\cite{ANS84} are reproduced remarkably well. Focusing on multiplier
distributions, we will now reveal limitations of log-stable
weight distributions.

Concentrating first on a discrete implementation of multiplicative cascade 
processes, where according to (\ref{two1})-(\ref{two3}) the parameters
are set as $\lambda^\prime=2$, $n=10$ and 
$0\leq x \leq 10^5L$, the associated log-stable weight distribution $p(q)$
is constructed by noting that $-\ln q \sim S_\alpha(\sigma,\beta,\mu)$ is
distributed according to a stable distribution \cite{SAM94}. The skewness
parameter has to be set equal to $\beta=1$ and the scale and shift
parameters $\sigma$ and $\mu$ are fixed by the scaling exponents
$\tau(1)=0$ and $\tau(2)=0.22$. For various stable parameters 
$0 < \alpha \leq 2$ 
the respective log-stable weight distributions are illustrated in Fig.\
4a. Depending on $\alpha$ these distributions possess different asymmetries
with respect to the average $\langle q \rangle =1$; 
in fact the skewness 
$S_3 = \langle(q - \langle q \rangle)^3\rangle 
       / \langle(q - \langle q \rangle)^2\rangle^{1.5}$, 
which should neither be confused with the skewness parameter
$\beta$ nor with the notation $S_\alpha$ of the stable distributions,
amounts to $S_3 = -0.43$, $0.06$, $0.63$, $1.28$ for
$\alpha = 1.25$, $1.5$, $1.75$, $2.00$. If a correct skewness is really 
a good criterion to reproduce unconditional as well as conditional 
multiplier distributions, then according to what we now know from the 
previous Sections it should be around $S_3 \approx 1$ for 
$\lambda^\prime=2$. This would limit the stable index $\alpha$ to a
regime close to $2$.

The unconditional left-sided base-two multiplier distributions resulting 
from the log-stable weight distributions with a given index $\alpha$ are
again scale-independent in the scale range $8\eta^* \leq l \leq L/2$
and are illustrated in Fig.\ 4b.
Since the log-stable weight distributions have already been fitted 
to the scaling exponent $\tau(2)$, 
there is no further free parameter left to adjust to the observed 
unconditional left-sided multiplier distribution (\ref{three2}) with 
$\beta_1=\beta_2=3.2$. For $\alpha \geq 1.5$ the agreement is acceptable. 
Notice especially the extra contributions at $M=0$ and $M=1$
for small values of $\alpha$. This is a consequence of
the algebraic tails of the stable-distributions, which lead to excess 
contributions at $q \approx 0$ in the corresponding weight distributions;
consult again Fig.\ 4a. 

The conditional multiplier distributions 
$p(M(l;2,\pm 0.5)|\underline{M}\leq M(2l;2,\pm 0.5)\leq \overline{M})$ are
also shown in Fig.\ 4b, where over all four possible left/right-sided
daughter/parent multiplier combinations has been averaged. For $\alpha=2$
a conditioning on a small parent multiplier, i.e.\ 
$(\underline{M},\overline{M})=(0,0.5)$, leads to a distribution more narrow
than the unconditional distribution and a conditioning on a large parent
multiplier, i.e.\ 
$(\underline{M},\overline{M})=(0.5,1)$, leads to a broader distribution. 
This is in qualitative agreement with the experimental findings of Ref.\
\cite{SRE95}. For $\alpha=1.75$ this tendency has almost vanished, for 
$\alpha=1.5$ it has vanished and for $\alpha=1.25$ it has even been reversed.
A very similar behaviour is observed in the conditional centred base-two
multiplier distributions, which are depicted in Fig.\ 4c; only for
$\alpha$ very close to 2 they are in qualitative agreement with the 
experimental findings of Ref.\ \cite{PED96}. This outcome also confirms
our conjecture from above that a correct positive skewness is needed for
the weight distribution in order to reproduce the multiplier distributions.  

A quasi-continuous implementation of multiplicative cascade processes with 
log-stable weight distributions leads to multiplier distributions, which are
identical to those shown in Figs.\ 4b+c for the discrete implementation. 
Hence, we arrive at the conclusion that although scaling exponents are
well fitted by log-stable weight distributions with index 
$\alpha\approx 1.5$ \cite{KID91,SCH92b} the observed systematics of the
multiplier distributions rules out such a stable index value. 
The log-normal limit $\alpha\approx 2$ reproduces the multiplier 
distributions much better.

\section{Conclusions}

Discrete and continuous implementations of geometric multiplicative 
cascade processes are both able to model multifractal characteristics
observed in the surrogate energy dissipation field of fully developed
turbulence. Since their relation to the Navier-Stokes equation remains
unclear neither form of implementation is principally favoured over 
the other. Although perhaps somewhat unexpected, this 
indistinguishability remains once a specific class of observables,
going beyond scaling exponents, is employed: multiplier distributions
referring to various scale-steps, discrete or quasi-continuous, do not 
care whether they result from discrete or continuous versions of
nonconservative multiplicative cascade processes. The reason for this 
is that the multivariate model implementation implies a small-scale
resummation leading to scale-independent fix-point distributions.
From an analysis point of view a discrete or continuous
multiplicative cascade process
with no correlations between the various branchings appears as a
continuous multiplicative cascade process with correlations.

There is more to learn from these fix-point multiplier distributions:
not every weight distribution associated to the cascade generator, 
which matches scaling exponents within experimental error bars, 
qualifies to reproduce the observed multiplier correlations. The weight
distribution has to possess a positive skewness. For example, 
so-called log-stable distributions with a stable index not close to
$\alpha=2$ do not share this property and are unable to reproduce the
observed conditional multiplier distributions. 
--
This leaves us with a speculation:
The original idea to directly access scaling exponents via unconditional
multiplier distributions does not work due to the observed correlations
between multipliers. However, these correlations help to restrict the
broad class of cascade weight distributions, which all reproduce the 
observed lowest-order scaling exponents, to those having a positive
skewness. A further restriction of this subclass appears to be possible
with yet additional observables to be derived from the full analytic 
solution of the multivariate cascade characteristic function 
\cite{GRE98}. Then, by directly accessing the cascade weight distribution,
it is tempting to study its dependence on the Reynolds number and 
on the flow configuration, and, thus, to check for universality.

\acknowledgements
B.J.\ acknowledges support from the Alexander-von-Humboldt Stiftung.

\newpage

\newpage
\begin{figure}
\begin{centering}
\epsfig{file=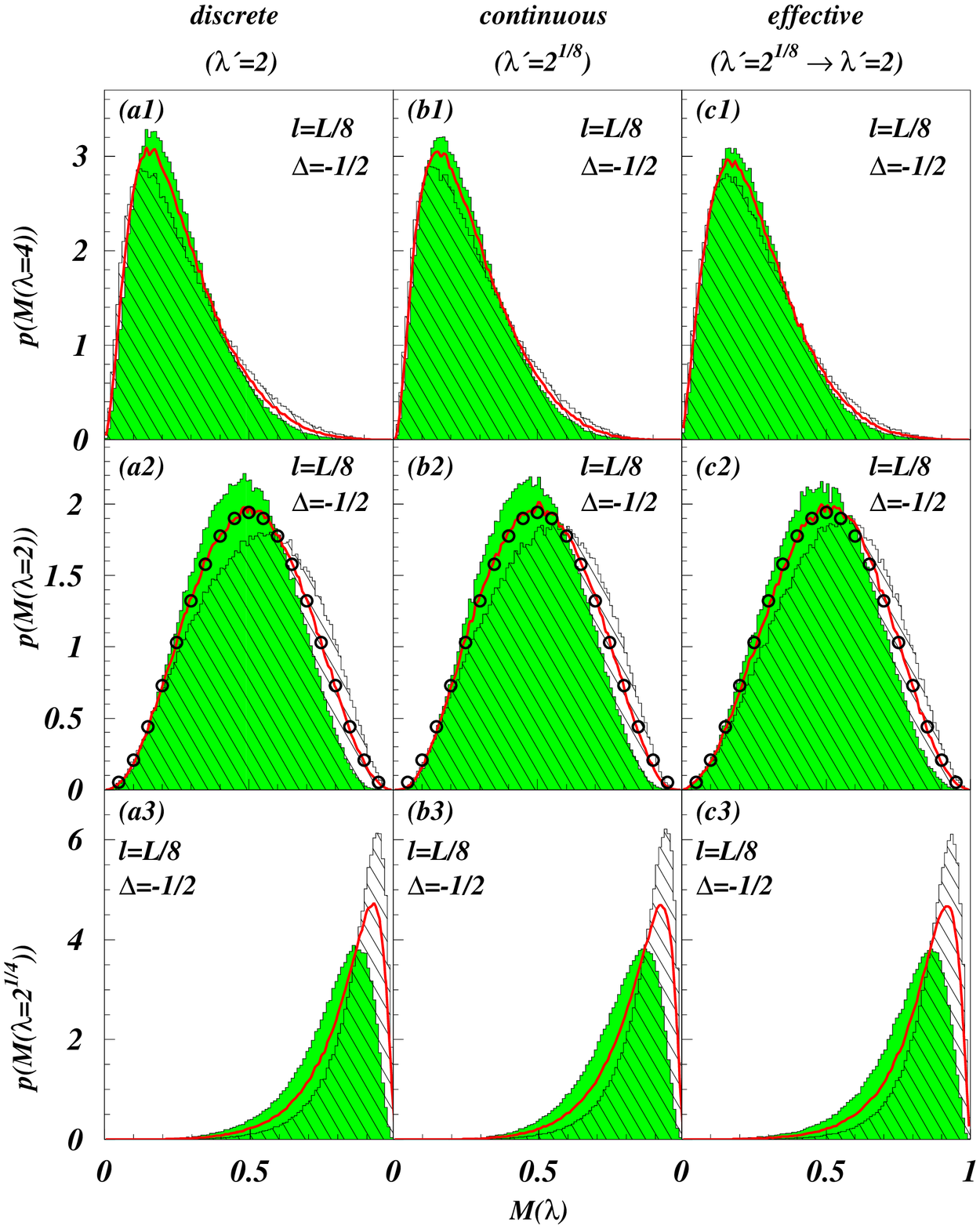,width=18cm}
\caption{
Scale-independent left-sided ($\Delta=-0.5$) multiplier distributions 
from a discrete (a1-3), a continuous (b1-3), 
and an effective, continuous-turned-discrete (c1-3) implementation 
of multiplicative cascade models; see text for model details.
Solid curve: unconditional distribution $p(M(l;\lambda,-0.5)$;
fully underlaid curve: conditional distribution
$p(M(l;\lambda,-0.5) | 0 \leq M(\lambda l;\lambda,-0.5) \leq \lambda^{-1})$;
hatched curve: conditional distribution
$p(M(l;\lambda,-0.5) | \lambda^{-1} \leq M(\lambda l;\lambda,-0.5) \leq 1)$.
The circled curve in (a2-c2) represents the parametrisation (\ref{three2})
with $\beta_1=\beta_2=3.2$.
} 
\end{centering}
\end{figure}

\newpage
\begin{figure}
\begin{centering}
\epsfig{file=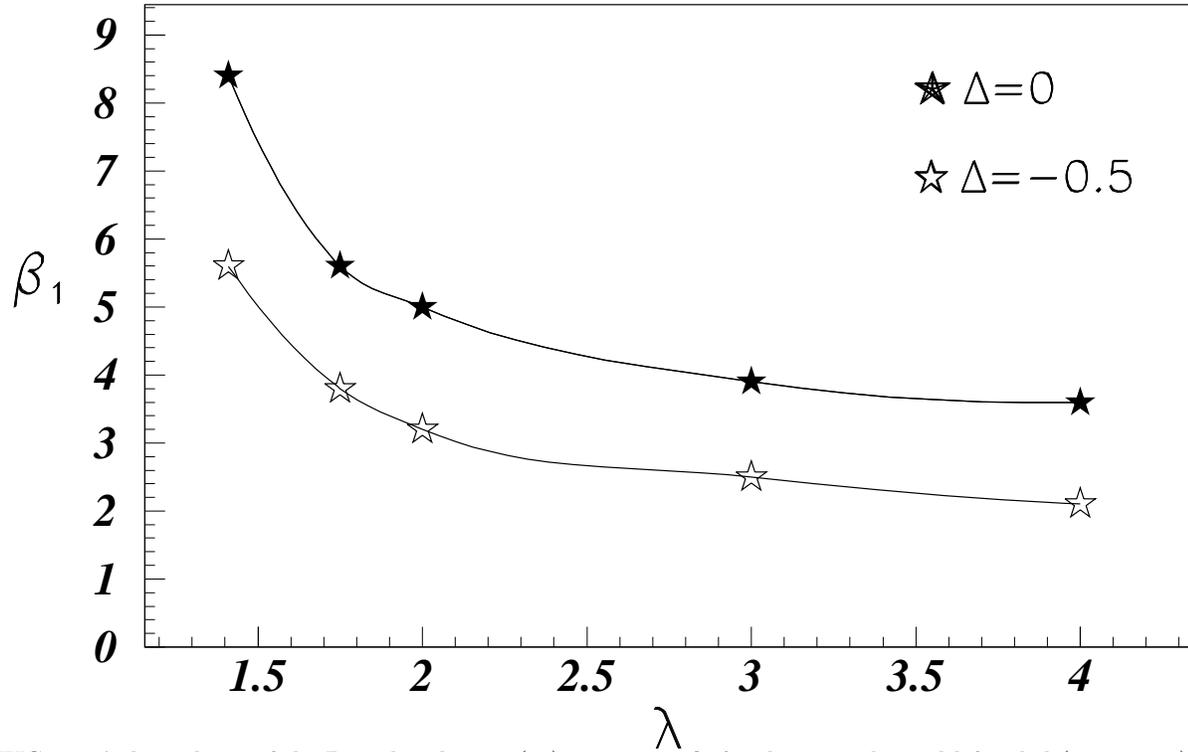,width=18cm}
\caption{
$\lambda$-dependence of the Beta-distribution-(\ref{three2})
parameter $\beta_1$ for the unconditional left-sided (open stars) and 
centred (full stars) multiplier distributions resulting from the
binomial weight distribution (\ref{three1}) with parameters
$\alpha_1=0.3$, $\alpha_2=0.65$. The marked points have been
calculated from least-square fits and the interpolating lines
are to guide the eye.
} 
\end{centering}
\end{figure}

\newpage
\begin{figure}
\begin{centering}
\epsfig{file=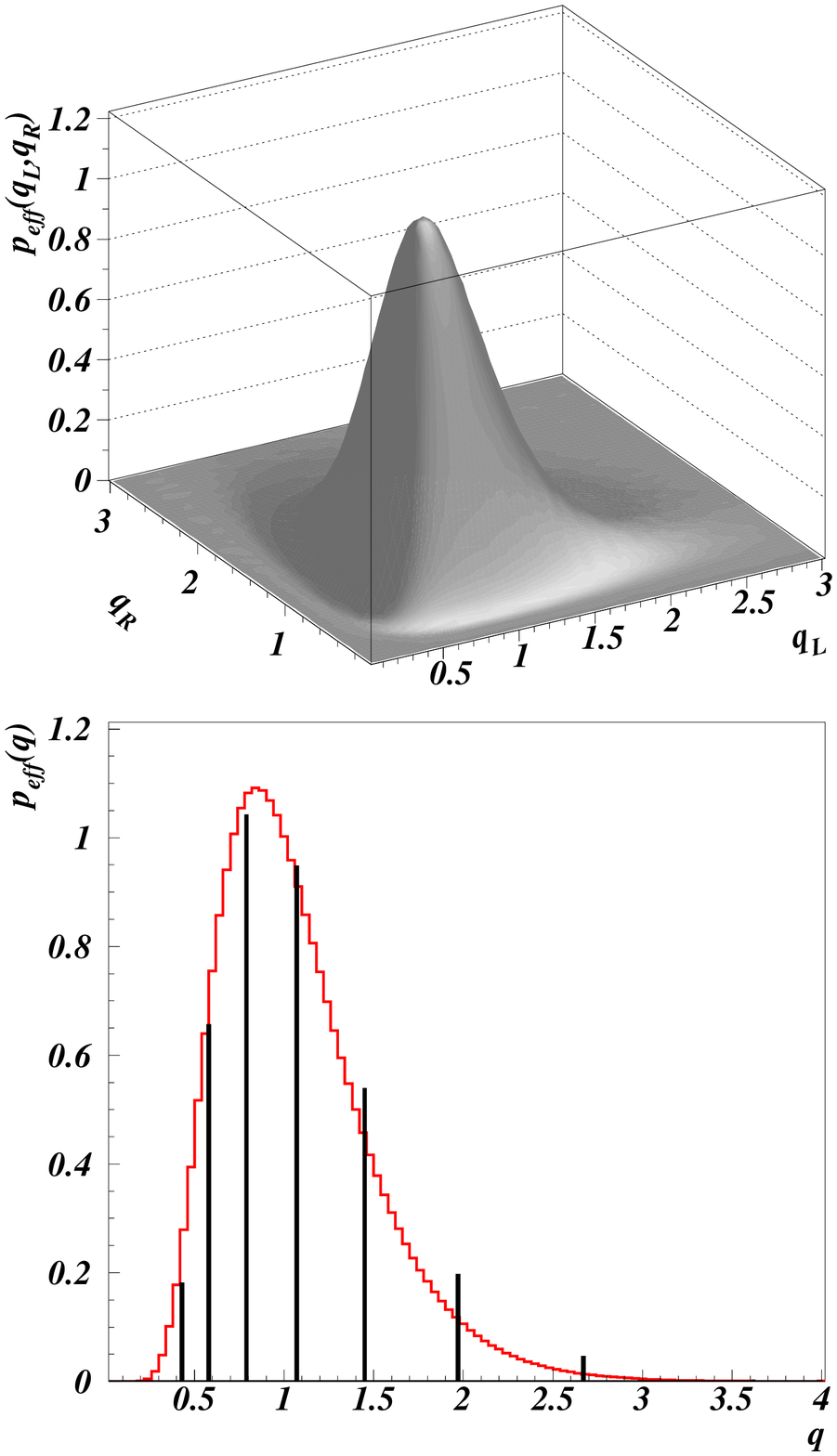,width=12cm}
\caption{
Effective bivariate (a) and projected (b) splitting function resulting
from a continuous multiplicative cascade process translated into a 
binary discrete one. For comparison the weight distribution resulting from
a naive $d$-fold convolution of the infinitesimal cascade generator is shown
in (b) as appropriately rescaled bars.
} 
\end{centering}
\end{figure}

\newpage
\begin{figure}
\begin{centering}
\epsfig{file=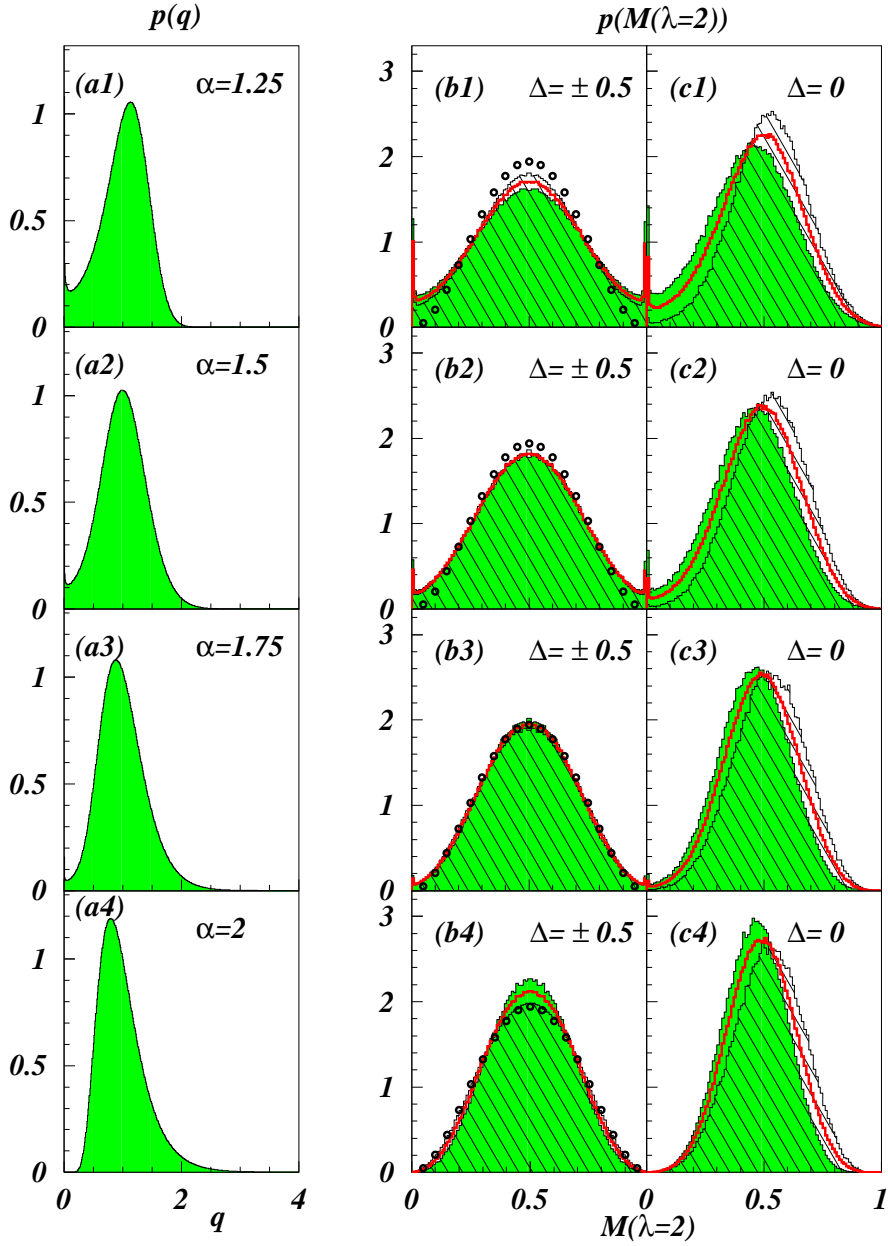,width=15cm}
\caption{
Various log-stable weight distributions (a) referring to a binary discrete
multiplicative cascade process and their resulting left/right-sided (b)
and centred (c) multiplier distributions.
} 
\end{centering}
\end{figure}

\end{document}